\documentclass[10pts,onecolumn]{elsart}
\usepackage{graphics}
\usepackage{amssymb}
\begin{document}
\begin{frontmatter}
  
  \title{Fast energy transfer mediated by multi-quanta bound states in
    a nonlinear quantum lattice}

  \author{C. Falvo and V. Pouthier} \address{CNRS UMR6624,
    Universit\'{e} de Franche Comt\'{e}, Laboratoire de Physique
    Mol\'{e}culaire, 25030 Besan\c {c}on cedex - France}

  \author{J. C. Eilbeck} \address{Department of Mathematics, Heriot-Watt
    University, Riccarton, Edinburgh, EH14 4AS - UK}

\begin{abstract}
  By using a Generalized Hubbard model for bosons, the energy transfer
  in a nonlinear quantum lattice is studied, with special emphasis on
  the interplay between local and nonlocal nonlinearity. For a
  strong local nonlinearity, it is shown that the creation of $v$
  quanta on one site excites a soliton band formed by bound states
  involving $v$ quanta trapped on the same site. The energy is first
  localized on the excited site over a significant timescale and then
  slowly delocalizes along the lattice. As when increasing the
  nonlocal nonlinearity, a faster dynamics occurs and the energy
  propagates more rapidly along the lattice. Nevertheless, the larger
  is the number of quanta, the slower is the dynamics. However, it is
  shown that when the nonlocal nonlinearity reaches a critical value,
  the lattice suddenly supports a very fast energy propagation whose
  dynamics is almost independent on the number of quanta.  The energy
  is transfered by specific bound states formed by the superimposition
  of states involving $v-p$ quanta trapped on one site and $p$ quanta
  trapped on the nearest neighbour sites, with $p=0,..,v-1$. These
  bound states behave as independent quanta and they exhibit a
  dynamics which is insensitive to the nonlinearity and controlled by the
  single quantum hopping constant.
\end{abstract}

\end{frontmatter}

\section{Introduction}

The concept of energy localization due to nonlinearity in classical
lattices has been a central topic of intense research over the
last three decades. This topic can be traced back to the seminal
works of Davydov, where the nonlinearity was introduced to explain the
vibrational energy flow in $\alpha$-helices \cite{kn:davydov}. The
main idea was that the energy released by the hydrolysis of adenosine
triphophate (ATP), partially stored in the high-frequency amide-I
vibration of a peptide group, delocalizes along the helix leading to
the formation of vibrational excitons. Due to their interaction with
the phonons of the helix, the excitons experience a nonlinear
dynamics. Therefore, they propagate according as a ``Davydov'' soliton,
a solution of the Nonlinear Schr\"{o}dinger (NLS) equation
within the continuum approximation \cite{kn:scott,kn:chris}. In the
mid-1980's, lattice effects were introduced through the analysis of
the discrete version of NLS. This equation supports specific solutions called ``lattice solitons'' \cite{kn:eilbeck1} and it has revealed the occurrence of a remarkable feature known as the self-trapping mechanism
\cite{kn:eilbeck11}. As discovered by Sievers and Takeno
\cite{kn:takeno}, the self-trapping is a special example of a more
general solutions called discrete breathers
\cite{kn:aubry,kn:flach,kn:mackay}. In classical anharmonic lattices,
a discrete breathers correspond to time-periodic and spatially localized
solutions which result from the interplay between the discreteness and
the nonlinearity. These solutions do not require integrability for
their existence and stability and it has been suggested that they
should correspond to quite general and robust solutions. Since
discrete breathers sustain a local accumulation of the vibrational
energy, which might be pinned in the lattice or may travel through it,
they are expected to be of fundamental importance.

At present, because the occurrence of classical breathers is a
relatively well understood phenomena, great attention has been paid to
characterize their quantum equivalent, for which less detailed results
are known \cite{kn:fleurov}. In the quantum regime, the Bloch theorem
applies due to the translational invariance of the lattice. Therefore,
the corresponding eigenstates cannot localize the energy because they
must share the symmetry of the translation operator which commutes
with the lattice Hamiltonian. Nevertheless, the nonlinearity is
responsible for the occurrence of specific states called multi-quanta
bound states
\cite{kn:fleurov,kn:kimball,kn:bogani,kn:eilbeck2,kn:eilbeck3,kn:eilbeck4,kn:bernstein,kn:eilbeck5,kn:dorignac,kn:proville,kn:pouthier1,kn:pouthier2,kn:pouthier3,kn:pouthier4,kn:pouthier5,kn:pouthier6,kn:pouthier7,kn:pouthier8}.
A bound state corresponds to the trapping of several quanta over only
a few neighbouring sites, with a resulting energy which is less than
the energy of quanta lying far apart. The distance separating 
the quanta is small, so that they behave as a single particle
delocalized along the lattice with a well-defined momentum. Since the
occurrence of these bound states results from the nonlinearity, they can be
viewed as the quantum counterpart of breathers or solitons.  In
low-dimensional lattices, two-quanta bound states have been observed
in molecular adsorbates such as H/Si(111) \cite{kn:guyot1,kn:sih},
H/C(111) \cite{kn:shen}, CO/NaCl(100) \cite{kn:dai} and CO/Ru(001)
\cite{kn:jakob1,kn:jakob2,kn:jakob3,kn:jakob4,kn:jakob5} using optical
probes. Bound states in the system H/Ni(111) were investigated via
high resolution electron energy loss spectroscopy \cite{kn:okuyama}.
Moreover, a recent experiment, based on femtosecond infrared
pump-probe spectroscopy, has clearly established the existence of
bound states in $\alpha$-helices \cite{kn:pouthier9}.

From a theoretical point of view, most of the previous work was
performed within the quantum equivalent of the discrete NLS equation.
The corresponding Hamiltonian is essentially a Bose version of the
Hubbard model which has been used to study a great variety of
situations ranging from molecular lattices \cite{kn:kimball} to
Bose-Einstein condensates \cite{kn:bishop1,kn:bishop2,kn:bishop3}.
Within this model, the nonlinearity is local so that it is responsible
for a strong interaction between quanta located on the same site.
However, through the small polaron description of the vibrational
energy flow in proteins, it has been shown recently that a nonlocal
nonlinearity strongly modifies the nature of the bound states
\cite{kn:pouthier3,kn:pouthier4,kn:pouthier5,kn:pouthier6,kn:pouthier10}.
In such systems, the polaron results from the dressing of a
vibrational exciton by a virtual cloud of phonons. In addition to an
attractive coupling between polarons located onto the same site, a
coupling takes place when polarons lie on nearest neighbour sites, due
to the overlap between their virtual cloud of phonons. When two quanta
are excited, the competition between the local and the nonlocal
nonlinearity favours the occurrence of two kinds of bound states whose
properties have been discussed in details in Ref. \cite{kn:pouthier3}.

At this step, the fundamental question arises whether the interplay
between the local and the nonlocal nonlinearity modifies the dynamics
of bound states involving several quanta. This is the purpose of the
present paper.  It will be shown, using a generalized Hubbard
model, that a critical value of the nonlocal nonlinearity favours a
resonance responsible for a fast energy transfer. Note that, although
such an approach is rather general, it will be applied to vibrational
excitons moving in a nonlinear molecular lattice.

The paper is organized as follows. In Section 2, the generalized
Hubbard model is described and the number state method
\cite{kn:eilbeck5}, used to define its eigenstates, is summarized.
Then, the quantum dynamics required to characterize the transport
properties is introduced. The corresponding time dependent
Schr\"{o}dinger equation is solved numerically in Section 3 where a
detailed analysis of the multi-quanta dynamics is performed. Finally,
the numerical results are discussed and interpreted in Section 4.

\section{Model Hamiltonian and Quantum analysis}

\subsection{Hamiltonian and Quantum states}

We consider a one-dimensional lattice formed by $N$ sites whose
position is defined in terms of the integer index $n$. Each site $n$
is occupied by a high-frequency oscillator described by the standard
boson operators $b^{\dag}_{n}$ and $b_{n}$. The lattice dynamics is
governed by a generalized Hubbard model for bosons whose Hamiltonian
is written as (using the convention $\hbar=1$)
\begin{equation}
  H=\sum_{n}
  \omega_{0}b^{\dag}_{n}b_{n} -Ab^{\dag}_{n}b^{\dag}_{n}b_{n}b_{n}
  -Bb^{\dag}_{n+1}b^{\dag}_{n}b_{n+1}b_{n} +\Phi[b^{\dag}_{n}b_{n+1} 
   +b^{\dag}_{n+1}b_{n}]
\label{eq:H}
\end{equation}
where $\omega_{0}$ is the internal frequency of each oscillator,
$\Phi$ is the hopping constant between nearest neighbour sites, and
$A$ and $B$ represent the local and the nonlocal nonlinearity,
respectively. Note that for vibrational excitons, the nonlinear
parameters are positive.

Since the Hamiltonian $H$ (Eq.(\ref{eq:H})) conserves the number of
quanta, its eigenstates can be determined by using the number state
method detailed in Ref. \cite{kn:eilbeck5}. This method is summarized
as follows.  Because $H$ commutes with the operator $\sum_{n}
b_{n}^{\dag}b_{n}$ which counts the total number of quanta, the
Hilbert space $E$ is written as the tensor product $E= E_{0} \otimes
E_{1} \otimes E_{2} \otimes ... \otimes E_{v} ...$, where $E_{v}$
refers to the $v$-quanta subspace. To generate $E_{v}$, a useful basis
set is formed by the local vectors $| p_{1},...,p_{N}\rangle$
($\sum_{n}p_n=v$), where $p_{n}$ is the number of quanta located on
the $n$th site. The dimension $d_v$ of $E_v$ is equal to the number of
ways of distributing $v$ indistinguishable quanta among $N$ sites, i.e.
$d_v=(N+v-1)!/v!(N-1)!$.

Within this representation, the Hamiltonian is block-diagonal.
However, the size of each block can be reduced by taking advantage of
the lattice periodicity. Indeed, due to the translational invariance,
the lattice momentum $k$ is a good quantum number so that an
``momentum'' basis set can be formed. For each $k$ value, an
momentum vector $| \Phi_{\alpha}(k) \rangle$ is obtained by
superimposing all the vectors $| p_{1},...,p_{N}\rangle$ describing
similar quanta distributions related to each other by a translation
along the lattice. For instance, the momentum vector built from
the $N$ vectors $| v,0,...,0\rangle$, ... , $| 0,0,...,v\rangle$ is
defined as
\begin{equation}
|  \Phi_1(k)\rangle = \frac{1}{\sqrt{N}} \sum_{n=0}^{N-1} e^{ikn}
T^{n}|  v,0,...,0\rangle
\label{eq:Phi1}
\end{equation}
where the translation operator $T$ satisfies $T|
p_{1},p_2...,p_{N}\rangle = |  p_{N},p_{1}...,p_{N-1}\rangle$.   

Consequently, in each subspace $E_v$, the Hamiltonian exhibits
independent blocks associated to each $k$ value and the corresponding
Schr\"{o}dinger equation can be solved with a minimum of computational
effort. This numerical procedure yields the eigenvalues
$\omega^{(v)}_{\lambda}(k)$, which define the dispersion relations of
the $v$-quanta subspace, and the associated eigenvectors
$\Psi^{(v)}_{\lambda}(k)$.

\subsection{Quantum dynamics}

The quantum dynamics is governed by the time dependent Schr\"{o}dinger
equation written as
\begin{equation}
i\frac{d|  \Psi (t) \rangle}{dt} =H |  \Psi (t) \rangle
\label{eq:schrod}
\end{equation}
where $| \Psi (t) \rangle$ is the lattice quantum state at time $t$.
By assuming that the state is known at the initial time $t=0$, its
value at time $t>0$ is expressed in terms of the eigenstates of $H$ as
\begin{equation}
  |  \Psi(t)\rangle =\sum_{v} \sum_{k \lambda}
  \langle\Psi^{(v)}_{\lambda}(k) 
  |  \Psi(0) \rangle e^{-i\omega^{(v)}_{\lambda}(k)t}|  
  \Psi^{(v)}_{\lambda}(k) \rangle
\label{eq:psit}
\end{equation}
The time dependent quantum state $| \Psi(t)\rangle$ is the central
object of the present study. Its knowledge allows us to characterize
the multi-quanta dynamics through the computation of the expectation
value of any relevant observable $O$ as $\langle O(t) \rangle =
\langle \Psi(t) | O | \Psi(t) \rangle$.  Nevertheless, the solution of
Eq.(\ref{eq:psit}) requires the specification of an initial quantum state. To
proceed, we restrict our attention to a spatially localized state
corresponding to the creation of $v$ quanta on the $n_0$th site as
\begin{equation}
|  \Psi(0)\rangle =\frac{b_{n_0}^{\dag v}}{\sqrt{v!}} |  0\rangle
\label{eq:psi0}
\end{equation}
where $| 0\rangle$ denotes the vacuum with zero quanta.  This specific
choice allows us to characterize the ability of the nonlinear quantum
lattice to localize the energy, at least over a given timescale.
Indeed, for a vanishing nonlocal nonlinearity ($B=0$), each subspace
$E_{v}$ supports a low energy band called the soliton band
\cite{kn:bernstein,kn:eilbeck5}. For vibrational excitons, the nonlinearity $A$ is
usually greater than the hopping constant $\Phi$. Consequently, the
soliton band describes $v$-quanta bound states in which the $v$ quanta
are trapped on the same site and behave as a single particle. When $v$
quanta are created on a given state, only the soliton band is
significantly excited. Because the dispersion of the band scales as
$\Phi^{v}/A^{v-1}$ when $A\gg \Phi$, the $v$ quanta are localized in
the vicinity of the excited site over a timescale which increases with
both the local nonlinearity and the number of quanta. This localized
behaviour, which is the quantum signature of the classical
self-trapping, disappears in the long time limit due to the non
vanishing dispersion of the soliton band.

In that context, the aim of the present paper is to analyze the way
this scenario is modified when the nonlocal nonlinearity is turned on.
 
\section{Numerical results}

In this section, the numerical diagonalization of the Hamiltonian $H$
is performed and the time dependent Schr\"{o}dinger equation is
solved. To realize the simulation, the hopping constant $\Phi$ is used
as a reduced unit and the local nonlinearity is fixed to $A=3\Phi$.
The nonlocal nonlinearity is taken as a free positive parameter.

\begin{figure*}
\begin{center}
\begin{minipage}[c]{0.68\linewidth}
\includegraphics{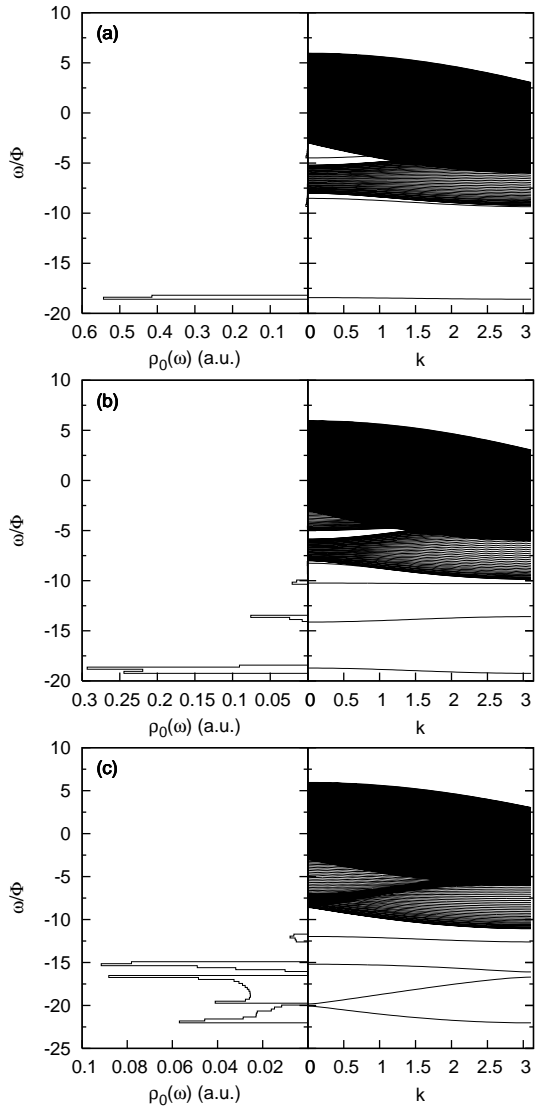}
\end{minipage}\hfill
\begin{minipage}[c]{0.32\linewidth}
  \caption{Energy spectrum for $N=91$, $v=3$ and $A=3\Phi$ and for (a)
    $B=0$, (b) $B=3\Phi$ and (c) $B=6\Phi$. The left-hand side of the
    figure represents the corresponding local density of state (see
    the text).}
\end{minipage}
\end{center}
\end{figure*}

When $v=3$, the energy spectrum of the Hamiltonian $H$ is displayed in
Fig. 1 for $B=0$ (Fig. 1a), $B=3\Phi$ (Fig. 1b) and $B=6\Phi$ (Fig.
1c). In each figure, the spectrum is centred onto the frequency
$3\omega_{0}$ and it corresponds to the dispersion curves drawn in
half of the first Brillouin zone of the lattice, i.e. $0<k<\pi$. The
left-hand side of Fig. 1 represents the corresponding local density of
state (LDOS). It describes the weight of the initial state $| \Psi(0)
\rangle$ in the eigenstates the energy of which ranges between
$\omega$ and $\omega +d\omega$ as
\begin{equation}
\rho_0(\omega)=\sum_{k \lambda} |  \langle\Psi^{(3)}_{\lambda}(k) |  
\Psi(0) \rangle | ^{2} \delta(\omega-\omega^{(3)}_{\lambda}(k))
\label{eq:LDOS0}
\end{equation}

When $B=0$ (Fig. 1a), the energy spectrum exhibits two energy continua
and three isolated bands. The high frequency continuum characterizes
free states describing three independent quanta. The low frequency
continuum supports states in which two quanta are trapped on the same
site whereas the third quantum propagates independently.  By contrast,
the isolated bands refers to bound states. The low frequency band is
the well known soliton band. It lies below the continua over the
entire Brillouin zone and it characterizes three quanta trapped on the
same site and delocalized along the lattice. The two other bands,
which lie respectively below and above the low frequency continuum,
describe bound states in which two quanta are trapped on a given site,
whereas the third quantum is trapped onto the corresponding nearest
neighbour sites. As shown in the left-hand side of Fig. 1a, the LDOS is
strongly peaked in the frequency range of the soliton band, which is
significantly excited when three quanta are initially created on the
same site.

When $B=3\Phi$ (Fig. 1b), the energy spectrum supports three continua
and four isolated bands. Indeed, a third continuum occurs between the
two previous energy continua. It describes states for which two quanta
are trapped on two nearest neighbour sites whereas the third quantum
propagates independently. The low frequency isolated band is the
soliton band, whose width has been increased when compared with the
previous situation. Above the soliton band, the next two isolated
bands correspond to the previously observed bound states in which two
quanta are trapped on one site whereas the third quantum is localized
on nearest neighbour sites. Note that these bands have been strongly
redshifted. Finally, a fourth bound state band occurs just below the
continua. The analysis of the corresponding eigenstates reveals that
this band supports bound states involving three quanta trapped on
three nearest neighbour sites. In addition to the occurrence of this
latter band, the behaviour of the LDOS reveals that the non vanishing
$B$ value is responsible for a weak hybridization between the three
lowest bound state bands. In other words, although the soliton band
still mainly refers to three quanta on the same site, the two other
bound states involve such a configuration but to a lesser extent.
 
Finally, when $B=6\Phi$ (Fig. 1c), the energy spectrum still supports
three continua and four isolated bands. The shape of the continua has
been slightly modified and the high frequency bound state band, which
still mainly refers to three quanta trapped onto three neighbouring
sites, has been slightly redshifted. However, strong modifications
affect the behaviour of the three low energy bound state bands. Indeed,
the gaps between these bands are now very small, which indicates the
occurrence of a strong hybridization. The LDOS exhibits a significant
value over the frequency range of the three low energy bands. It shows
four peaks for specific frequencies, corresponding either to the bottom
or to the top of the three low frequency bands (see Fig. 1c). In other
words, these bands support bound states which are superpositions of
states involving three quanta trapped on one site and two quanta on
one site trapped with a third quantum localized on neighbouring sites.

\begin{figure}
\includegraphics{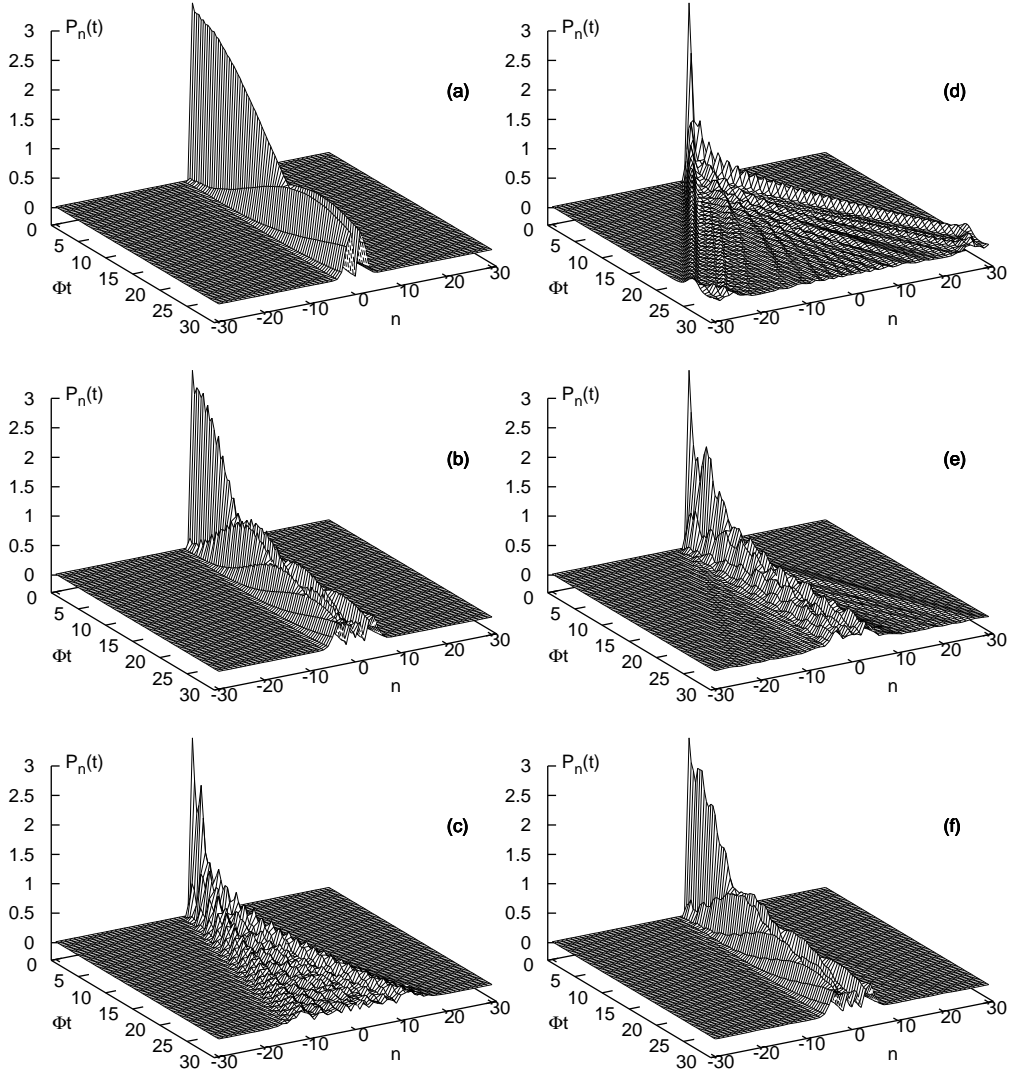}
\caption{Space and time evolution of the number of quanta for $N=91$
  and $v=3$ and for (a) $B=0$, (b) $B=2\Phi$, (c) $B=4\Phi$, (d)
  $B=6\Phi$, (e) $B=8\Phi$ and (f) $B=10\Phi$}
\end{figure}

In Fig. 2, the time evolution of the expectation value of the
population operator, $P_n(t)=\langle \Psi(t)| b_nb^{\dag}_n | \Psi(t)
\rangle$, is displayed for $v=3$ and for different $B$ values. The
size of the lattice if fixed to $N=91$ sites. When $B=0$ (Fig. 2a),
the population of the initial site $P_{n_{0}}(t)$ decreases very
slowly with time. It reaches $50$ $\%$ of its initial value when $t$
is about $13.5\Phi^{-1}$, which indicates that the energy stays
localized on the excited site $n_0$ over a significant timescale.
Nevertheless, energy propagation occurs and the population delocalizes
along the lattice. However, this propagation corresponds to a rather
slow process, since 95 \% of the popuplation is concentrated in the 6
sites surrounding $n_0$ at time $t=30\Phi^{-1}$. When $B=2\Phi$ (Fig.
2b), the same features occur. Nevertheless, a faster dynamics takes
place since $P_{n_{0}}(t)$ reaches $50$ $\%$ of its initial value for
$t=6.8\Phi^{-1}$. The energy propagates slowly and 95 \% of the
population is concentrated in the 10 sites surrounding $n_0$ for
$t=30\Phi^{-1}$. When $B=4\Phi$ (Fig. 2c), the population of the
excited site still decays but it now supports high frequency
oscillations. Its decay is enhanced, and it reaches $50$ $\%$ of its
initial value when $t=2.1\Phi^{-1}$. The population propagates more
easily along the lattice and its maximum value is observed on the
sites $n_0\pm12$ for $t=30\Phi^{-1}$. In that case, 95 \% of the
population is contained in the 30 sites surrounding $n_0$.

\begin{figure*}
\begin{center}
\begin{minipage}[c]{0.68\linewidth}
\includegraphics{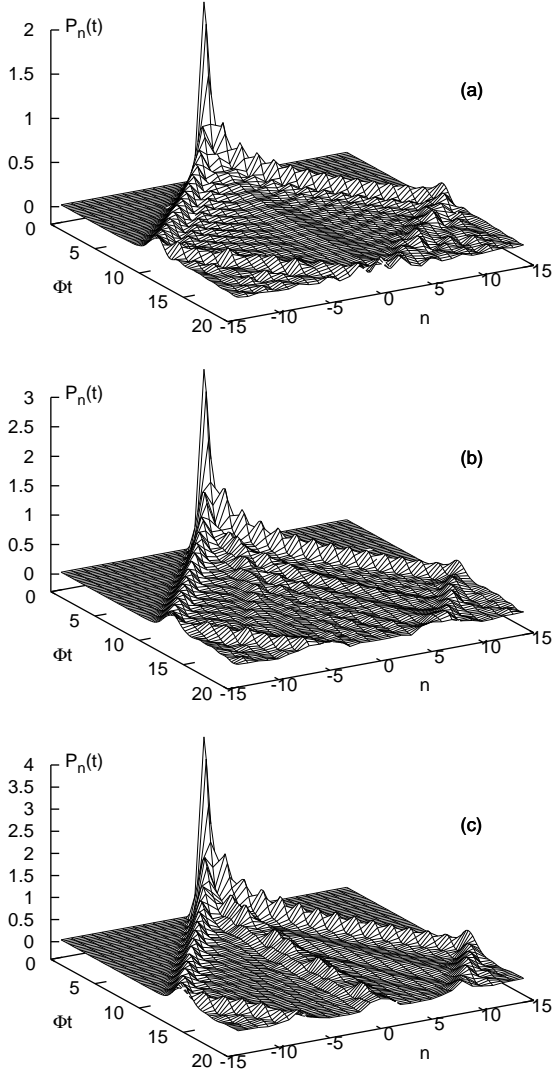}
\end{minipage}\hfill
\begin{minipage}[c]{0.32\linewidth}
  \caption{Space and time evolution of the number of quanta for $N=29$
    and $B=2A$ and for (a) $v=2$, (b) $v=3$ and (c) $v=4$.}
\end{minipage}
\end{center}
\end{figure*}

When $B=6\Phi$ (Fig. 2d), a fully different behaviour is observed.
Indeed, $P_{n_{0}}(t)$ shows a very fast decay since it reaches 50 \%
of its initial value for $t=0.6\Phi^{-1}$. This decay occurs over a
timescale of about 20 times shorter than the time associated to the
decay of the population for $B=0$. In addition, Fig. 2d clearly shows
that the lattice supports a rather fast energy propagation. Two
population wave packets are emitted on each side of the initial site
with a velocity of about $\Phi$. In other words, the maximum value of
the population is observed on the sites $n_{0}\pm35$ for
$t=30\Phi^{-1}$.

Finally, for greater $B$ values, a rather slow dynamics recurs. When
$B=8\Phi$ (Fig. 2e) and $B=10\Phi$ (Fig. 2f), the population of the
excited site slowly decreases. It supports a small amplitude
high-frequency modulation, and it reaches 50 \% of its initial value
for $t=4.2\Phi^{-1}$ and $t=6.3\Phi^{-1}$ when $B=8\Phi$ and
$B=10\Phi$, respectively. The population propagates along the lattice
with a rather small velocity. Therefore, when $t=30\Phi^{-1}$, 95 \%
of the energy is concentrated on the 18 sites surrounding the excited
site when $B=8\Phi$. The size of this region decreases to 14 sites
when $B=10\Phi$.

Fig. 2 has clearly shown the occurrence of a specific dynamics when
the relation $B=2A$ is verified. As illustrated in Fig.3, such a
behaviour does not depend significantly on the value of the number of
quanta. Indeed, the figure shows the time evolution of the population
for $v=2$ (Fig. 3a), $v=3$ (Fig. 3b) and $v=4$ (Fig. 3c) and for
$B=2A$. Note that the lattice size is fixed to $N=29$ to avoid an
overlong computational time, especially when $v=4$. Whatever the
number of quanta, the population of the excited site decays over a
short timescale. Then, two population wave packets are emitted on each
side of the excited site with a velocity which slightly decreases with
the number of quanta. When $v=2$, the wave packets reach the edges of
the lattice for $t=10.8\Phi^{-1}$ whereas longer delays
$t=12.6\Phi^{-1}$ and $t=13.7\Phi^{-1}$ are required when $v=3$ and
$v=4$, respectively.

The time evolution of the survival probability is displayed in Fig. 4
for $v=2$ (Fig. 4a), $v=3$ (Fig. 4b) and $v=4$ (Fig. 4c). The number
of sites is fixed to $N=29$ and three different $B$ values have been
used, i.e. $B=0$ (full line), $B=3\Phi$ (dashed line) and $B=6\Phi$
(dotted line). The survival probability $S_0(t)$ is defined as the
probability to observe the lattice in the state $| \Psi(0) \rangle$ at
time $t$, i.e. $S_0(t)=| \langle \Psi(0) | \Psi(t) \rangle | ^{2}$. It
is expressed in terms of the Fourier transform of the LDOS
(Eq.(\ref{eq:LDOS0})) as
\begin{equation}
S_0(t)=|  \int_{-\infty}^{+\infty} \rho_0(\omega)e^{-i\omega t}d\omega |  ^{2}
\label{eq:s1}
\end{equation}

When $B=0$, the survival probability slowly decreases with time, which
indicates that the energy is localized on the excited site over a
significant timescale which increases with $v$. For instance, the
first zero of $S_0(t)$ is reached for $t=4\Phi^{-1}$ and
$t=30\Phi^{-1}$ when $v=2$ and $3$, respectively. When $v=4$, $S_0(t)$
is almost constant over the timescale displayed in the figure. Note
that $S_0(t)$ supports a small amplitude high-frequency modulation
which is clearly seen for $v=3$ and $v=4$.  As when increasing $B$,
$S_0(t)$ exhibits a faster decay. When $B=3\Phi$, the first zero of
the survival probability occurs for $t=2.4\Phi^{-1}$ and
$t=9.4\Phi^{-1}$ when $v=2$ and $3$, respectively. In a similar way,
although the first zero cannot be observed for $v=4$, $S_0(t)$
decreases with time (Fig. 4c). In a marked contrast with the case
$B=0$, the amplitude of the modulation increases with $B$.  Finally,
when $B=6\Phi$, $S_0(t)$ suddenly shows a very fast decay whatever the
number of quanta. This decay occurs over a similar timescale and a
first minimum is reached for $t=0.9\Phi^{-1}$, $t=0.75\Phi^{-1}$ and
$t=0.7\Phi^{-1}$ when $v=2$, $3$ and $4$, respectively. In addition,
due to the small lattice size, a revival is observed for
$t=20\Phi^{-1}$, $t=25\Phi^{-1}$ and $t=27\Phi^{-1}$ when $v=2$, $3$
and $4$, respectively. These latter times represent the delay for the
excitation to cover the whole lattice, so that the corresponding
velocities are about $1.44\Phi$, $1.16\Phi$ and $1.06\Phi$. Therefore,
in a perfect agreement with the results observed in Fig. 3, this
feature indicates that when $B=2A$ a fast energy propagation occurs.
The initial excitation covers the lattice over a timescale almost
independent on the number of quanta and with a velocity typically of
the order of the velocity of a single quantum insensitive to any
nonlinearity.
 
\begin{figure*}
\begin{center}
\begin{minipage}[c]{0.68\linewidth}
\includegraphics{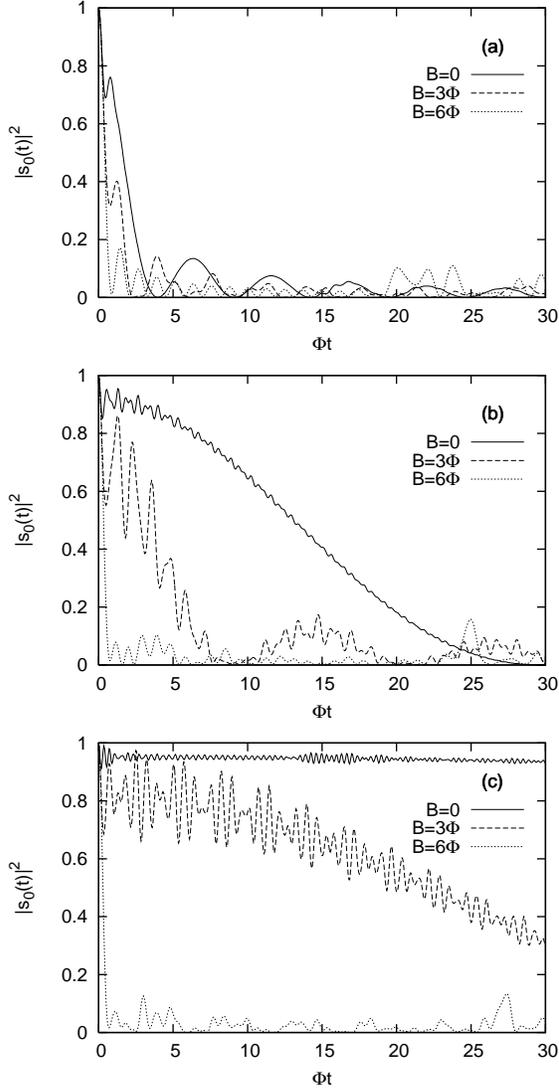}
\end{minipage}\hfill
\begin{minipage}[c]{0.32\linewidth}
  \caption{Survival probability for $N=29$ and for (a) $v=2$, (b)
    $v=3$ and (c) $v=4$. Three $B$ values have been used, i.e. $B=0$
    (full line), $B=3\Phi$ (dashed line) and $B=6\Phi$ (dotted line).}
\end{minipage}
\end{center}
\end{figure*}

\begin{figure*}
\begin{center}
\begin{minipage}[c]{0.68\linewidth}
\includegraphics{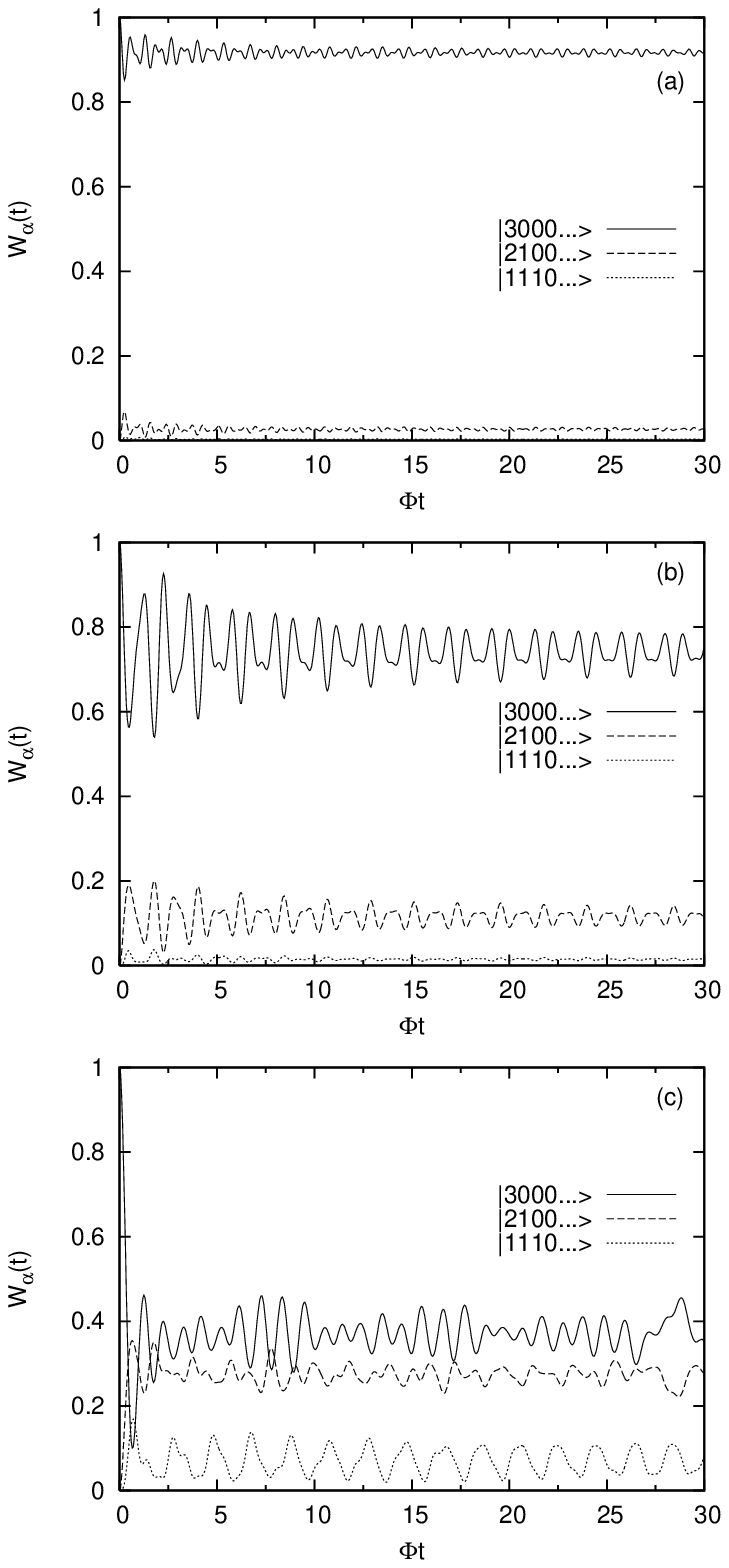}
\end{minipage}\hfill
\begin{minipage}[c]{0.32\linewidth}
  \caption{Relevant number states involved in the quantum dynamics for
    (a) $B=0$, (b) $B=3\Phi$ and (c) $B=6\Phi$.}
\end{minipage}
\end{center}
\end{figure*}

To determine the relevant number states which contribute significantly
to the quantum dynamics, we introduce $W_{\alpha}(t)$ as the sum
over $k$ of the weight of each intermediate basis vector $|
\Phi_{\alpha}(k) \rangle$ as
\begin{equation}
W_{\alpha}(t)=\sum_{k} |  \langle \Phi_{\alpha}(k)  |  \Psi(t)\rangle | ^{2}
\end{equation}
Since a number state refers to a given distribution of the quanta, the
associated momentum vector characterizes all the local states
connected to that distribution but related to each other through a
translation along the lattice. Therefore, the sum over $k$ allows us
to characterize the participation of such a distribution in the
quantum dynamics. For instance, from the $N$ local vectors $|
v,0,...,0\rangle$, ... , $| 0,0,...,v\rangle$, we have defined in
Eq.(\ref{eq:Phi1}) the  vector $| \Phi_1(k)\rangle$ which
basically describes $v$ quanta trapped on the same site. The
corresponding weight $W_{1}(t)$ allows us to quantify the participation of such a
configuration.

\begin{figure*}
\begin{center}
\begin{minipage}[c]{0.68\linewidth}
\includegraphics{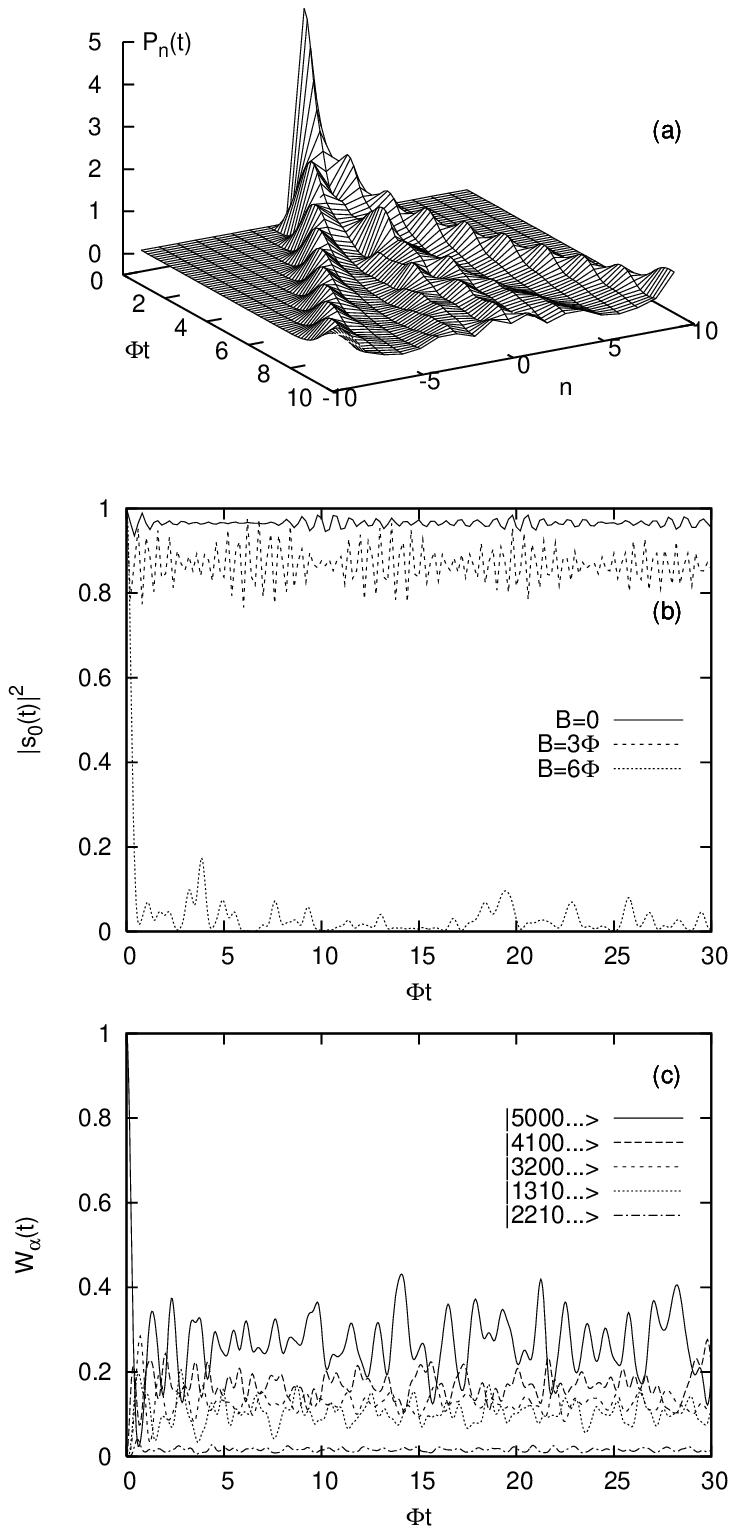}
\end{minipage}\hfill
\begin{minipage}[c]{0.32\linewidth}
  \caption{Lattice dynamics for $v=5$ and $N=19$. (a) Space and time
    evolution of the number of quanta, (b) survival probability and
    (c) relevant number states.}
\end{minipage}
\end{center}
\end{figure*}

For $v=3$, Fig.\ 5 displays the relevant number states involved in the
quantum dynamics for $B=0$ (Fig. 5a), $B=3\Phi$ (Fig.\ 5b) and
$B=6\Phi$ (Fig.\ 5c). When $B=0$, the states involving $3$ quanta
trapped on the same site control more than 90 \% of the quantum
dynamics. However, as when increasing $B$, the participation of other
states takes place. When $B=3\Phi$ (Fig.\ 5b), the states describing
$3$ quanta trapped on the same site remain the most relevant since
they carry about 75 \% of the full dynamics. However, about 22 \% of
the dynamics is controlled by states involving two quanta on one site
trapped to a third quantum located on the nearest neighbour site. A
factor 2 is included, due to the participation of the two equivalent
states $| 2100...\rangle$ and $| 1200...\rangle$. Finally, when
$B=6\Phi$ (Fig. 5c), the dynamics are mainly controlled by the states
referring to two quanta on one site trapped to a third quantum. These
states, including the factor two, carry about 56 \% of the full
dynamics whereas 35 \% originates in states involving $3$ quanta
trapped on one site.  Note that a third type of states connected to
three quanta trapped onto three nearest neighbour sites participate to
a lesser extent, since they share only 8 \% of the dynamics.

Finally, the different observed features are illustrated in Fig.\ 6 for
$v=5$. The lattice size is fixed to $N=19$ to avoid an overlong
computational time.  Nevertheless, Fig.\ 6a shows the occurrence of the
delocalization of the population when $B=2A$. As when $v=2$, $3$, and
$4$, a fast decay of the population of the excited site takes place.
It is followed by the emission of two wave packets which propagate
according to velocity of about $\Phi$. In Fig.\ 6b, the survival
probability indicates a strong energy localization when $B=0$ and
$B=3\Phi$, whereas it decreases rapidly over a timescale of about
$0.5\Phi^{-1}$ when $B=2A$. Finally, Fig.\ 6c shows that the main part
of the quantum dynamics is described by the states involving the
configurations $| 500...\rangle$ (about 26 \%), $| 410...\rangle$ and
$| 140...\rangle$ (about 40 \%), $| 320...\rangle$ and $|
230...\rangle$ (about 22 \%) and $| 131...\rangle$ (about 8 \%).

\section{Interpretation and discussion}

In the previous section, the numerical results have revealed a strong
dependence of the lattice dynamics on the nonlocal nonlinearity $B$.
Indeed, for a vanishing $B$ value, the creation of $v$ quanta on one
site mainly excites the soliton band of the energy spectrum. This
excitation gives rise to the occurrence of the quantum equivalent of
the classical self-trapping phenomena. The energy is localized on the
excited site over a significant timescale, which increases with both
the local nonlinearity and the number of quanta. Then, a rather slow
energy flow takes place due to the finite value of the soliton
bandwidth. As when increasing the nonlocal nonlinearity, a faster
dynamics occurs. The survival probability, which characterizes the
memory of the initial state in the time dependent quantum state,
decreases more rapidly and supports a significant high frequency
modulation. Similarly, the population propagates more rapidly along
the lattice. In that regime, the speed of the dynamics still depends
on the number of quanta and the larger is the number of quanta, the
slower is the dynamics.  However, when the nonlocal nonlinearity
reaches the critical value $B=2A$, a very different behaviour takes
place.  Indeed, the lattice supports a very fast energy propagation
whose dynamics is almost independent of the number of quanta. The
survival probability decreases over a timescale of about $\Phi^{-1}$,
and two population wave packets emitted on each side of the excited
site propagate with a velocity typically of about $\Phi$. Our
numerical results have clearly established that this fast energy
transfer is mediated by bound states. However, the timescale of the
dynamics is of the same order of magnitude as the timescale
governing the free state dynamics. In other words, the bound states
involved in the dynamics when $B=2A$ behave like independent quanta
insensitive to the nonlinearity.  Finally, when $B$ exceeds its
critical value, the quantum self-trapping regime recurs.

As shown in Section 3, the main part of the dynamics is controlled by
specific number states. Indeed, we have verified numerically for
$v=2,...,7$ that the most relevant states involve number states of the
form $| v-p,p,0,...,0\rangle$, with $p=0,..,v-1$. Note that the states
$| 1,v-2,1,...,0\rangle$, $| 1,v-3,2,...,0\rangle$ and $|
2,v-3,1,...,0\rangle$ participate to a lesser extend in the dynamics.
Therefore, the relevant states, which describe $v-p$ quanta on one
site trapped with $p$ quanta on a nearest neighbour site, are
characterized by the self-energies
$\epsilon_{p}=v\omega_{0}-v(v-1)A+p(v-p)(2A-B)$ (see Eq.(\ref{eq:H})).
As long as $| 2A-B | $ is strong enough, the different relevant
configurations are weakly coupled to each other. Therefore only the
number states $| v,0,0,...,0\rangle$ participate significantly to the
dynamics which follows the initial creation of $v$ quanta on one site.
However, as illustrated in Fig. 5, a strong hybridization between the
different configurations occurs when $| 2A-B | $ tends to zero.
Finally, when $B=2A$, a resonance takes place since all the relevant
states have the same energy.

To understand more clearly the way this resonance modifies the energy
transfer in the nonlinear lattice, we can take advantage of the fact
that only relevant number states participate in the full dynamics.
This feature allows us to establish a simplified model, which is able
to account for the dynamics in the case $A\gg \Phi$. This model is
introduced in the next section.

\subsection{Equivalent lattice model}

As mentioned previously, a good description of the dynamics in the
subspace $E_{v}$ is obtained by restricting the number state basis to
the set of the relevant vectors. This set is formed by the $N\times v$
vectors defined as
\begin{equation}
|  n,p\rangle = |  p_1=0,p_2=0,...,p_n=v-p,p_{n+1}=p,... 0 \rangle
\label{eq:vec}
\end{equation} 
where $| n,p\rangle$ characterizes $v-p$ quanta on the $n$th site and
$p$ quanta of the site $n+1$. The representation of the Hamiltonian
$H$ in the restricted basis is equivalent to a tight-binding model on
the lattice displayed in Fig. 7a. To simplify the notation, the
restriction of the full Hamiltonian will still be denoted $H$. Within
this model, each site supports the state $| n,p\rangle$ whose energy
is $\epsilon_{p}=p(v-p)\epsilon$, where $\epsilon=2A-B$. Note that
$\epsilon_0=0$ has been used as the origin of the energy. Nearest
neighbour sites are coupled to each other through generalized hopping
constants. The hopping constant between $| n,p\rangle$ and $|
n,p+1\rangle$ is equal to $\Phi_{p}=\sqrt{(p+1)(v-p)}\Phi$, with
$p=0,..,v-2$. By symmetry, the hopping constant between $|
n,v-1\rangle$ and $| n+1,0\rangle$ is $\Phi_{0}=\sqrt{v}\Phi$.

\begin{figure*}
\begin{center}
\begin{minipage}[c]{0.68\linewidth}
\includegraphics{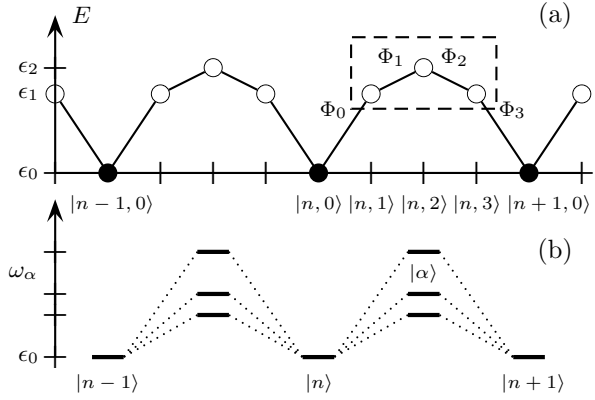}
\end{minipage}\hfill
\begin{minipage}[c]{0.32\linewidth}
\caption{Equivalent lattice model.}
\end{minipage}
\end{center}
\end{figure*}

Consequently, the dynamics of the quantum nonlinear lattice is
formally equivalent to the dynamics of a fictitious particle moving on
the lattice shown in Fig. 7a. At time $t=0$, the particle is created
in the state $| n_0,0\rangle$ and its time evolution is governed by
the corresponding Schr\"{o}dinger equation. Unfortunately, although
the model has been greatly simplified, this equation cannot be solved
analytically. However, the knowledge of some general properties is
sufficient to clarify our understanding of the physics involved in the
energy transfer. In addition, the model allows us to determine the
LDOS which characterizes the weight of a localized state $| n,0
\rangle$ in the eigenstates of the lattice. It represents a measure of
the response of the lattice to the initial excitation and gives the
survival probability simply by performing a Fourier transform.

\subsection{General properties of the lattice eigenstates}

The lattice displayed in Fig. 7a exhibits a translational invariance
and its unit cell is defined in terms of two subsystems. The first
subsystem is formed by the single state $| n,0 \rangle$, whereas the
second subsystem involves the $v-1$ coupled states $| n,p \rangle$,
with $p=1,..,v-1$. A moment's reflection will convince the reader that
this latter subsystem represents a modified nonlinear quantum dimer.
More precisely, it is the representation of a dimer in the number
state basis in which the two states involving $v$ quanta on the first
site and $v$ quanta on the second site have been suppressed. As for
the dimer, this subsystem is invariant under the reflection symmetry
which transforms the state $| n,p \rangle$ into the state $| n,v-p
\rangle$.

Due to the translational invariance, the lattice wave vector $k$ is a
good quantum number. Therefore, for each $k$ value, the quantum states
are described by a $v \times v$ matrix $H(k)$. Instead of representing
this matrix in the local basis $| n,p \rangle$, it is more convenient
to introduce a new basis set formed by the single state $| n,0
\rangle$ and by the $v-1$ eigenstates of the modified dimer. Indeed,
although the quantum states of a modified dimer are only known for
small $v$ values, they represent a key ingredient needed to understand
the physics of the lattice. Therefore, let $| \alpha \rangle$ denotes
the $\alpha$th eigenstate of a modified dimer whose energy is
$\omega_{\alpha}$ (see Fig. 7b). Because of the reflection symmetry,
each eigenstate is either symmetric or antisymmetric. As a result, the
wave function $\psi_{\alpha p}=\langle n,p | \alpha \rangle$ satisfies
$\psi_{\alpha p}=\pm \psi_{\alpha v-p}$, depending on whether the
state is symmetric ($+$) or antisymmetric ($-$). For even $v$ values,
there are $N_S=v/2$ symmetric and $N_A=(v-2)/2$ antisymmetric
eigenstates. By contrast, for odd $v$ values, $N_S=N_A=(v-1)/2$.

In that context, the matrix Hamiltonian $H(k)$ is expressed as
\begin{equation}
H(k)= \left(
\begin{array}{cccc} 
   0 & x_{\alpha}(k) & x_{\alpha'}(k) & ...\\
   x^{*}_{\alpha}(k) & \omega_{\alpha} & 0 & ...\\
   x^{*}_{\alpha'}(k) & 0 & \omega_{\alpha'} & ...\\
   ... & ... & .... & ...
\end{array}
\right)
\label{eq:Hk}
\end{equation}
where $x_{\alpha}(k)=\sqrt{v}\phi ( \psi_{\alpha 1}+\psi_{\alpha v-1}
\exp(-ik))$ is the coupling between the state $| n,0 \rangle$ and $|
\alpha \rangle$ which results from the interaction between $| n,0
\rangle$ with both the state $| n,1 \rangle$ of the same unit cell and
the state $| n-1, v-1 \rangle$ of the previous unit cell (see Fig. 7).

>From Eq.(\ref{eq:Hk}), the delocalization of the fictitious particle
originates in the coupling between two neighbouring states $| n,0
\rangle$ and $| n+1,0 \rangle$, through their interaction with the
eigenstates of the $n$th modified dimer. As a consequence, each
lattice eigenstate is a superimposition involving the localized states
and the eigenstates of a modified dimer. However, such a
hybridization depends both on the energy difference $\omega_{\alpha}$,
and on the strength of the coupling $x_{\alpha}(k)$, which is
drastically sensitive to the symmetry of the dimer eigenstates.

Eq.(\ref{eq:Hk}) allows us to determine the exact expression of the LDOS.
Indeed, according to the standard definition used in condensed matter
physics, the LDOS is expressed in terms of the Green's operator
$G(k,\omega)=(\omega-H(k))^{-1}$ as
\begin{equation}
  \rho_{0}(\omega)=-\frac{1}{N\pi} Im \sum_{k} \langle n_0,0 |  
      G(k,\omega+i0^{+}) |  n_0,0\rangle
\label{eq:LDOS1}
\end{equation}
Since the LDOS only depends on the restriction of the Green's operator
to $| n_0,0\rangle$, it can be obtained by applying standard
projection methods. Therefore, by taking advantage of the symmetry of
the eigenstates $| \alpha \rangle$, the restricted Green's operator is
written as
\begin{equation}
  \langle n_0,0 |  G(k,\omega) |  n_0,0\rangle=\frac{1}{\omega 
 -\Delta(\omega)-2J(\omega)\cos(k)}
\label{eq:Gk}
\end{equation}
where 
\begin{eqnarray}
\Delta(\omega)&=&2v\Phi^{2}\sum_{\alpha}\frac{\psi_{\alpha
    1}\psi_{\alpha 1}}
{\omega-\omega_{\alpha}} \nonumber \\
J(\omega)&=& v\Phi^{2}\sum_{\alpha}\frac{\psi_{\alpha 1}\psi_{\alpha
    v-1}}
 {\omega-\omega_{\alpha}}
\label{eq:DJ}
\end{eqnarray}
By inserting Eq.(\ref{eq:Gk}) into Eq.(\ref{eq:LDOS1}), the LDOS is
finally written as
\begin{equation}
\rho_{0}(\omega)=\frac{1}{\pi}\frac{1}{\sqrt{(2J(\omega)+\Delta(\omega)
 -\omega)(2J(\omega)-\Delta(\omega)+\omega))}}
\label{eq:LDOSG}
\end{equation}
Eq.(\ref{eq:LDOSG}) allows us to establish two general properties.
First, the zeros of the LDOS are given by the poles of the parameters
$\Delta(\omega)$ and $J(\omega)$. In that context, Eq.(\ref{eq:DJ})
clearly shows that the LDOS supports $v-1$ zeros, which corresponds to
the eigenenergies of the modified dimer. In the vicinity of a zero,
the LDOS scales as $\sqrt{| \omega-\omega_{\alpha}| }$. Then, the LDOS
exhibits two kinds of poles which are given by the solutions of the
two equations $2J(\omega)\pm(\Delta(\omega)-\omega)=0$. From
Eq.(\ref{eq:DJ}), these two equations are rewritten as
\begin{eqnarray}
4v\Phi^{2}\sum^{S}_{\alpha}\frac{\psi^{2}_{\alpha1}}{\omega-\omega_{\alpha}}&
=&\omega \nonumber \\
4v\Phi^{2}\sum^{A}_{\alpha}\frac{\psi^{2}_{\alpha1}}{\omega-\omega_{\alpha}}&
=&\omega
\label{eq:pole}
\end{eqnarray}
where the symbol $\sum^{S,A}$ denotes a sum over either the symmetric
or the antisymmetric eigenstates of a modified dimer. In
Eq.(\ref{eq:pole}), the first equation gives $N_S+1$ solutions whereas
the second equation yields $N_A+1$ solutions. Therefore, the LDOS
supports $N_S+N_A+2=v+1$ poles. In the vicinity of a pole
$\omega_{po}$, the LDOS thus diverges according to the power law $|
\omega-\omega_{po}| ^{-1/2}$.

These results reveal the occurrence of zeros and poles in the LDOS, as
observed in Fig. 1. In addition, they point out the important role
of the symmetry of the eigenstates of the dimer, especially for
specific values of the lattice wave vector $k$. Indeed, when $k=0$,
the coupling $x_{\alpha}(k=0)$ between $| n,0 \rangle$ and each
antisymmetric state $| \alpha \rangle$ vanishes. As a consequence,
$H(k=0)$ (Eq.(\ref{eq:Hk})) has $N_A$ eigenvalues equal to the
eigenenergies of the antisymmetric states of the modified dimer. They
describe lattice eigenstates which do not involve $| n,0 \rangle$, so
 they corresponds to zeros of the LDOS. By contrast, the remaining
coupling between $| n,0 \rangle$ and the symmetric states $| \alpha
\rangle$ produces $N_S+1$ eigenstates which induce divergences in the
LDOS.  When $k=\pi$, the same features occur, but by inverting the role
played by the symmetric and the antisymmetric states.  The matrix
$H(k=\pi)$ exhibits $N_S$ eigenvalues equal to the eigenenergies of
the symmetric states $| \alpha \rangle$ and which correspond to zeros
of the LDOS. A strong coupling remains between $| n,0 \rangle$ and the
$N_A$ antisymmetric states, leading to $N_A+1$ eigenvalues which give
rise to divergences in the LDOS.

Finally, the coupling between $|n,0\rangle$ and the $v-1$ dimer
eigenstates $|\alpha \rangle$ is at the origin of the creation of the
lattice eigenstates. It is responsible for the occurrence of specific
signatures in the LDOS and control the lattice dynamics. However,
depending on the strength of the coupling, different dynamical
behaviours can be observed ranging from quantum self-trapping to fast
energy transfer. In the following of the text, these features are
first illustrated for the two simples situations corresponding to
$v=2$ and $v=3$. Then, a general discussion is given to interpret the
specific behaviour observed when $\epsilon=0$.

\subsection{Application to $v=2$ and $v=3$}

The case $v=2$ is rather simple, because a modified dimer involves a
single state $| n,1 \rangle$ whose energy is $\epsilon_{1}=\epsilon$.
Therefore, $H(k)$ is a $2\times 2$ matrix which describes the
hybridization between $| n,0 \rangle$ and $| n,1 \rangle$. It can be
solved exactly so that the lattice supports two bands whose dispersion
relations are expressed as
\begin{equation}
  E_{\pm}(k)=\frac{\epsilon}{2}\pm\sqrt{(\frac{\epsilon}{2})^{2}+
 8\Phi^{2}\cos^{2}(k/2)}
\end{equation} 
The corresponding LDOS, which shows a single zero and three poles, is
defined as
\begin{equation}
\rho_{0}(\omega)=\frac{1}{\pi}\sqrt{\frac{-(\omega-\epsilon_{1})}
 {(\omega-E_{-}(0))(\omega-E_{-}(\pi))(\omega-E_{+}(0))}}
\label{eq:LDOS3}
\end{equation}

When $\epsilon\gg\Phi$ ($B \ll 2A$), the two bands are only weakly
coupled. The low frequency band, whose energy is about
$\epsilon_{0}=0$, refers to the delocalization of the fictitious
particle over the different states $| n,0\rangle$. By contrast, the
second band accounts for the delocalization of the fictitious particle
over the states $| n,1 \rangle$. In other words, the low frequency
band is the soliton band describing two quanta trapped on the same
site and delocalized along the lattice. The second band refers to the
delocalization of bound states involving two quanta trapped onto two
nearest neighbour sites.

Since only the soliton band is significantly excited, it controls the
dynamics. To understand this feature, let us evaluate the
corresponding survival probability (Eq.(\ref{eq:s1})). Although the
Fourier transform of Eq.(\ref{eq:LDOS3}) is not analytic in a general
way, it can be evaluated when the two bands lie far from each other.
In that case, the integral over the frequency range of a given band
can be performed by neglecting the frequency dependence of the other
band. After straightforward calculations, the survival probability is
thus expressed as
\begin{equation}
S_0(t) \approx J_0^{2}(\frac{4\Phi^{2}t}{\epsilon})[1-\frac{8\Phi^{2}}
 {\epsilon^{2}}(1-\cos(\epsilon t)) ]
\label{eq:S0V2}
\end{equation} 
where $J_0$ is the Bessel function of the first kind. The survival
probability slowly decreases with time and it reaches its first zero
for $t=2.4\epsilon/4\Phi^{2}$. Note that this value is extracted from
the first zero $t \approx 2.40$ of $J_0(t)$. When $A=3\Phi$ and $B=0$,
this value is equal to $3.6\Phi^{-1}$, in a perfect agreement with the
numerical results about $4\Phi^{-1}$ (see Fig. 4a). Eq.(\ref{eq:S0V2})
shows that the survival probability supports a high frequency
modulation whose amplitude decreases with $\epsilon$. Such a
modulation, whose frequency is equal to the energy difference between
$| n,0 \rangle$ and $| n,1 \rangle$, characterizes the small
participation of the second band in the dynamics.

As when $\epsilon$ decreases, i.e. when $B$ tends to $2A$, the width
of the low frequency band increases as a result of the redshift of its
low frequency edge. By contrast, the second band exhibits a strong
redshift accompanied by an increase in width. In addition, its
contribution to the LDOS increases, which indicates that the coupling
between the two kinds of bound states is enhanced. Such a behaviour
occurs until the high frequency edge of the soliton band meets the low
frequency edge of the second band. This feature takes place when
$\epsilon=0$, i.e. when $B=2A$. In that case, the LDOS displays a
single band, only, with two divergences whose frequencies are
$\pm2\sqrt{2}\Phi$. The corresponding bandwidth is thus a maximum and it
is equal to $4\sqrt{2}\Phi$. At the resonance, this band clearly
suggest strong hybridization between $| n,0\rangle$ and $| n,1
\rangle$. In that case, the Fourier transform of Eq.(\ref{eq:LDOS3})
can be determined exactly and it yields the following survival
probability
\begin{equation}
S_0(t)=J_0^{2}(2\sqrt{2}\Phi t)
\label{eq:S0V21}
\end{equation} 
The resonance induces a decay of the survival probability over a very
short timescale. It reaches its first zero for $t=0.85\Phi^{-1}$, in
perfect agreement with the results shown in Fig. 4a. This feature,
mainly due to the strong hybridization between the two bands, is
responsible for a fast energy transfer along the whole lattice. In
other words, the transport of energy results from a series of
transitions between states involving successively two quanta on the
same site and two quanta on nearest neighbour sites.

When $v=3$, a modified dimer involves the two states $| n,1\rangle$
and $| n,2 \rangle$, which have the same energy $2\epsilon$. The first
state describes two quanta on  site $n$ and one quantum on  site
$n+1$, whereas the second state refers to one quantum on  site $n$
and two quanta on  site $n+1$. Due to the coupling $\Phi_{1}=2\Phi$
between these states, a strong hybridization takes place so that the
eigenstates of the modified dimer correspond to a symmetric and to an
antisymmetric superimposition as
\begin{equation}
|  \alpha_{\pm} \rangle =\frac{1}{\sqrt{2}}(| n,1\rangle \pm | n,2 \rangle)
\label{eq:pm}
\end{equation}
The corresponding eigenenergies are $\omega_{\pm}=2\epsilon \pm
2\Phi$. Therefore, $H(k)$ is a $3\times 3$ matrix which describes the
hybridization between $| n,0 \rangle$ and $| \alpha_{\pm} \rangle$. In
a general way, the lattice supports three bands and the corresponding
LDOS is written as
\begin{equation}
\rho_{0}(\omega)=\frac{1}{\pi}\sqrt{\frac{-(\omega-\omega_+)(\omega-\omega_-)}
{(\omega-E_{1})(\omega-E_{2})(\omega-E_{3})(\omega-E_{4})}}
\label{eq:LDOS4}
\end{equation}
where the four poles are defined as 
\begin{eqnarray}
E_{1}&=&\epsilon-\Phi-\sqrt{\epsilon^2-2\epsilon \Phi+7\Phi^{2}} \nonumber \\
E_{2}&=&\epsilon+\Phi-\sqrt{\epsilon^2+2\epsilon \Phi+7\Phi^{2}} \nonumber \\
E_{3}&=&\epsilon-\Phi+\sqrt{\epsilon^2-2\epsilon \Phi+7\Phi^{2}} \nonumber \\
E_{4}&=&\epsilon+\Phi+\sqrt{\epsilon^2+2\epsilon \Phi+7\Phi^{2}}
\end{eqnarray}

\begin{figure*}
\begin{center}
\begin{minipage}[c]{0.68\linewidth}
\includegraphics{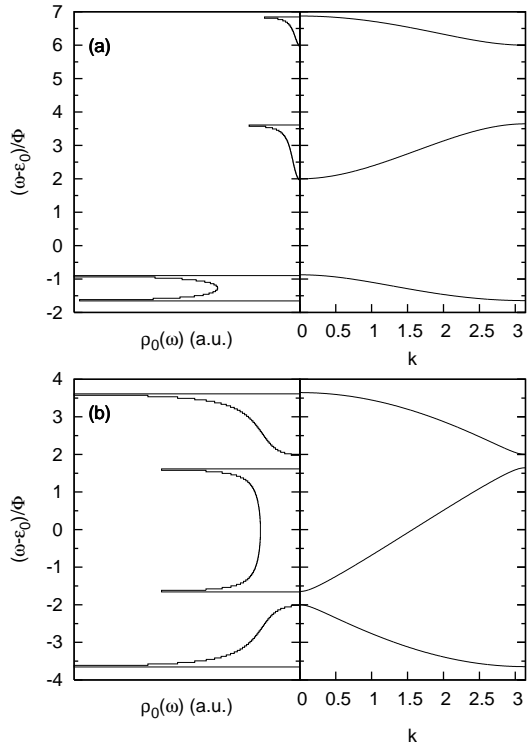}
\end{minipage}\hfill
\begin{minipage}[c]{0.32\linewidth}
  \caption{Theoretical energy spectrum and LDOS for $v=3$ and for (a)
    $\epsilon=2$ and (b) $\epsilon=0$.}
\end{minipage}
\end{center}
\end{figure*}

For strong $\epsilon$ values (see Fig. 8a), the three bands are well
separated, in a perfect agreement with the results displayed in Fig.
1a and 1b. As for $v=2$, the low frequency band mainly describes the
delocalization of the fictitious particle over the different states
$|n,0\rangle$. It thus corresponds to the soliton band, the LDOS of
which exhibits two divergences whose frequencies are $E_{1}$ and
$E_{2}$. The resulting bandwidth is thus about
$6\Phi^{3}/\epsilon^{2}$. By contrast, the two other bands account for
the delocalization of the fictitious particle over the eigenstates $|
\alpha_{\pm} \rangle$ belonging to different unit cells. Consequently,
the band just above the soliton band refers to bound states formed by
the antisymmetric superimposition of the states $| n,1 \rangle$ and $|
n,2 \rangle$, whereas the high frequency band characterizes bound
states involving their symmetric superimposition. As shown in Fig. 8a,
the LDOS connected to these two bands shows a single divergence and a
single zero.

Since only the soliton band is significantly excited when three quanta
are created on one site, it controls the main part of the dynamics.
The survival probability is thus approximately expressed as
\begin{eqnarray}
&&S_0(t) \approx
(1-\frac{3\Phi^{2}}{\epsilon^{2}})J_0^{2}(\frac{3\Phi^{3}t}
{\epsilon^{2}}) \nonumber \\
&+&\frac{3\Phi^{2}}{\epsilon^{2}}J_0(\frac{3\Phi^{3}t}{\epsilon^{2}})
[J_0(\frac{3\Phi^{2}t}{2\epsilon})\cos(2\epsilon t) - J_1(\frac{3\Phi^{2}t}
{2\epsilon})\sin(2\epsilon t)]\cos(2\Phi t)
\label{eq:S0V3}
\end{eqnarray}
The survival probability exhibits a rather slow dynamics governed by
the first term in the right-hand side of Eq.(\ref{eq:S0V3}). It
decreases over a timescale typically of about
$t=2.40\epsilon^{2}/3\Phi^{3}$. When $A=3\Phi$ and $B=0$, this time is
equal to $28.8\Phi^{-1}$, in a perfect agreement with the numerical
results $30\Phi^{-1}$ (see Fig. 4b). Note that the survival
probability supports a high frequency modulation whose amplitude
decreases with $\epsilon$. Such a modulation, whose frequency is
typically about $\omega_{\pm}$, accounts for the small hybridization
between the soliton band and the two other bands.

As for the case when $\epsilon$ decreases, i.e. when $B$ tends to
$2A$, the three bands are redshifted and their bandwidth increases. In
addition, the contribution to the LDOS of the two high frequency bands
is enhanced due to their hybridization with the soliton band.
Nevertheless, the three bands do not hybridize in a similar way.
Indeed, the soliton band hybridizes first with the antisymmetric band.
This mechanism dominates when $\epsilon=\Phi/4$, for which both the
soliton band and the antisymmetric band condense into a single band.
Then, for smaller $\epsilon$ values, the hybridization with the
symmetric band takes place. Finally, when $\epsilon=0$, the LDOS
exhibits three bands again (Fig. 8b). However, all these bands mix the
three different kinds of bound states so that the three bands control
the dynamics. As a result, the typical time governing the decay of the
survival probability is about $t=2\times2.4/\Delta\omega$, where
$\Delta\omega=2(1+\sqrt{7})\Phi$ is the difference between the high
frequency and the low frequency divergences of the LDOS. When
$A=3\Phi$ and $B=6\Phi$, this time is equal to $0.66\Phi^{-1}$ in
perfect agreement with the numerical results $0.75\Phi^{-1}$ (see Fig.
4b). As a consequence, a fast energy transfer takes place, with a
velocity of about the hopping constant $\Phi$, which characterizes the
motion of single quantum insensitive to the nonlinearity. However, as
for $v=2$, the transport of energy is mediated by states formed by the
superimposition of the three kinds of bound states.

\subsection{Interpretation of the resonance $\epsilon=0$}

The previous examples clearly illustrate the fact that the lattice
dynamics exhibits two distinct behaviours, depending on the value of the
parameter $\epsilon=2A-B$.

When $\epsilon\gg\Phi$ ($B\ll 2A$), the energy transfer is mediated by
bound states formed by $v$ quanta trapped on the same site and
behaving as a single particle. This particle delocalizes along the
lattice due to the coupling between nearest neighbour states
$|n,0\rangle$ through their interaction with the modified dimers.
Because the eigenenergies of the dimer lie far from the energy of the
localized state $|n,0\rangle$, this coupling is very weak.
Consequently, the $v$ quanta bound states belong to the low energy
soliton band, which is characterized by a very small bandwidth of about
$\Delta \omega \approx 4v\Phi^{v}/(v-1)!\epsilon^{v-1}$
\cite{kn:bernstein,kn:eilbeck5}. Therefore, the creation of $v$ quanta on one site
mainly excites the soliton band, leading to the occurrence of the
quantum equivalent of the classical self-trapping. The energy is
localized on the excited site over a significant timescale which
increases with both $\epsilon$ and $v$. It finally propagates slowly
at a very small velocity, proportional to the bandwidth.

By contrast, when $\epsilon=0$, i.e. when $B=2A$, a very different
behaviour takes place, mainly due to the occurrence of a resonance,
since all the states $|n,p \rangle$ have the same self-energy. This
resonance allows for a complete energy delocalization at a rather
large velocity, typically of about the hopping constant $\Phi$.  In a
marked contrast with the previous situation, the resonance induces a
strong interaction between the localized states and the modified
dimers. Nevertheless, as detailed below, the origin of this
interaction strongly depends on whether $v$ is even or odd, due to the
$v$ dependence of the energy spectrum of a modified dimer.

For even $v$ values, a modified dimer supports a single eigenstate,
denoted as $|\alpha_{0}\rangle$, whose energy vanishes. The $v-2$
remaining eigenstates are grouped into pairs formed by two eigenstates
having the same symmetry but with opposite energy. As a result, the
coupling between $|n,0\rangle$ and all these $v-2$ eigenstates appears
extremely weak. Indeed, by using a standard perturbation theory, it is
easy to show that the second order correction of the energy of
$|n,0\rangle$ due to these couplings is expressed as $\Delta
E=-\sum_{\alpha \neq \alpha_{0}} |x_{\alpha}(k)|^{2}/\omega_{\alpha}$.
This correction vanishes exactly because the sum involves pairs of
states which experience the same interaction with $|n,0\rangle$ but
which have an opposite energy. In other words, a pair of eigenstates
yields two pathways for the transition between $|n,0\rangle$ and
$|n+1,0\rangle$ which interfere destructively so that the resulting
probability amplitude vanishes.

Consequently, the relevant part of $H(k)$ reduces to a $2 \times 2$
matrix which describes the hybridization between $| n,0 \rangle$ and
the resonant eigenstate $|\alpha_{0}\rangle$. From Eq.(\ref{eq:Hk}),
it can be solved exactly so that the lattice supports two energy bands
whose dispersion relations are expressed as $\pm \sqrt{v} \Phi
|\psi_{\alpha_{0}1}+\psi_{\alpha_{0}v-1}\exp(-ik)|$. Depending on
whether $|\alpha_{0}\rangle$ is symmetric or antisymmetric, these two
bands vanish for $k=\pi$ or $k=0$. Therefore, the gap between these
two bands vanishes, so that the resulting bandwidth is equal to $\Delta
\omega = 4\sqrt{v}\Phi |\psi_{\alpha_{0}1}|$. By performing the
numerical diagonalization of the dimer Hamiltonian for $v$ ranging
from 2 to 200, it is straightforward to show that the wave function
behaves as $|\psi_{\alpha_{0}1}|\approx \sqrt{2/v}$. The bandwidth is
thus of about a few times the hopping constant as $\Delta
\omega\approx 4\sqrt{2}\Phi$. Note that the diagonalization of the
full $H(k)$ matrix Eq.(\ref{eq:Hk}) shows that the bandwidth slightly
decreases with the number of quanta. For instance it varies from
$5.66\Phi$ for $v=2$ to $4.50\Phi$ for $v=50$. This rather large
bandwidth is responsible for the decay of the survival probability
over a short timescale, which indicates the occurrence of a fast energy
transfer at a velocity typically of about the velocity of
independent quanta insensitive to the nonlinearity. Nevertheless, this
delocalization originates in the propagation of bound states, which
result from the hybridization between $v$-quanta trapped on the same
site and $v$-quanta distributed according to the resonant state
$|\alpha_{0} \rangle$. Our numerical analysis reveals that this
distribution involves the superimposition of the number states
$|v-1,1,0,0...\rangle$, $|v-3,3,0,0...\rangle$, $|v-5,5,0,0...\rangle$
...$|1,v-1,0,0...\rangle$.

For odd $v$ values, the modified dimer has $v-1$ eigenstates which are
grouped into pairs containing two states having an opposite energy and
a different symmetry. Therefore, since no resonance occurs with the
modified dimer, $|n,0 \rangle$ interacts with all the eigenstates
$|\alpha \rangle$. Nevertheless, among all these interactions, only
the coupling between $|n,0 \rangle$ and the two eigenstates whose
energy is close to zero is significant.  These two states, denoted
 by $|\alpha_{S} \rangle$ (symmetric) and $|\alpha_{A}
\rangle$ (antisymmetric) respectively, belong to the same group and have an
opposite energy $\omega_{S}=-\omega_{A}$. Consequently, the relevant
part of the $H(k)$ reduces to a $3 \times 3$ matrix written as
\begin{equation}
H(k)\approx \left(
\begin{array}{ccc} 
   0 & x_{S}(k) & x_{A}(k) \\
   x^{*}_{S}(k) & \omega_{S} & 0 \\
   x^{*}_{A}(k) & 0 & -\omega_{S} \\
\end{array}
\right)
\label{eq:Hk1}
\end{equation}
where $x_{S}(k)=\sqrt{v}\Phi \psi_{S 1}( 1+ \exp(-ik))$ and
$x_{A}(k)=\sqrt{v}\Phi \psi_{S 1}( 1- \exp(-ik))$.  As discussed in
Section 4.2, Eq.(\ref{eq:Hk1}) can be diagonalized for specific values
of the lattice wave vector $k$ by taking advantage of the dimer
eigenstate symmetry. Indeed, when $k=0$, the coupling between $| n,0
\rangle$ and $| \alpha_{A} \rangle$ vanishes. The energy $\omega_{A}$
is thus an eigenvalue of the matrix Hamiltonian Eq.(\ref{eq:Hk1})
which corresponds to a zero of the LDOS. However, the coupling between
$| n,0 \rangle$ and $| \alpha_{S} \rangle$ remains and it produces two
eigenvalues
$\omega_{S}/2\pm\sqrt{(\omega_{S}/2)^2+4v\Phi^2|\psi_{S1}|^2}$ which
correspond to two poles of the LDOS.  In the same way, when $k=\pi$,
there is no interaction between $|n,0\rangle$ and the symmetric state
so that $\omega_{S}$ is an eigenvalue of the Hamiltonian which is a
zero of the LDOS. By contrast, the coupling with the antisymmetric
state leads to two eigenvalues
$-\omega_{S}/2\pm\sqrt{(\omega_{S}/2)^2+4v\Phi^2|\psi_{S1}|^2}$ which
yield two poles of the LDOS.

Consequently, the energy spectrum supports three bands which describe
strong hybridizations between the three states $|n,0 \rangle$,
$|\alpha_{S} \rangle$ and $|\alpha_{A} \rangle$. These bands govern
the dynamics of the lattice which is controlled by a rather large
bandwidth defined as the energy difference between the high frequency
and the low frequency divergences of the LDOS as $\Delta
\omega=|\omega_{S}|+\sqrt{\omega_{S}^{2}+16v\Phi^{2}|\psi_{S1}|^2}$.
The numerical analysis reveals that this bandwidth is of about a few
times the hopping constant $\Phi$. However, as for even $v$ value, the
exact diagonalization of the full $H(k)$ matrix Eq.(\ref{eq:Hk})
reveals that the bandwidth slightly decreases with the number of
quanta. It varies from $7.29\Phi$ for $v=3$ to $6.72\Phi$ for $v=51$.
Therefore, this large bandwidth favours a fast decay of the survival
probability followed by a fast energy transfer. This energy transfer
is mediated by bound states formed by the superimposition of
$v$-quanta trapped on the same site and $v$-quanta distributed
according to the two states $|\alpha_{S} \rangle$ and $|\alpha_{A}
\rangle$ which involve all the number states $|v-1,1,0,0...\rangle$,
$|v-2,2,0,0...\rangle$, ...$|1,v-1,0,0...\rangle$.

\ack

The authors would like to acknowledge the " R\'{e}gion de Franche-Comté\'{e} "
which has partially supported this work through a BDI Grant.

\end{document}